\input harvmac
\input epsf.tex
\noblackbox
\def\pl#1#2#3{{Phys. Lett. } {#1}B (#2) #3}
\def\np#1#2#3{{Nucl. Phys. } B{#1} (#2) #3}

\def\prl#1#2#3{{Phys. Rev. Lett. } {#1} (#2) #3}
\def\physrev#1#2#3{{Phys. Rev. } {#1} (#2) #3}

\def\ie{{\it i.e.}}
\def\({\left(}
\def\){\right)}
\def\[{\left[}
\def\]{\right]}
\def\ltap{\ \raise.3ex\hbox{$<$\kern-.75em\lower1ex\hbox{$\sim$}}\ }
\def\gtap{\ \raise.3ex\hbox{$>$\kern-.75em\lower1ex\hbox{$\sim$}}\ }

\Title{ \vbox{
{\hfill NSF-ITP-94-67}\smallskip
{\hfill BUHEP-94-11}\smallskip
{\hfill DOE-ER-40561-146-INT94-0059}\smallskip
{\hfill UW/PT 9402}\smallskip}}
{\vbox{\centerline{Diffusion Enhances }
\vskip6pt\centerline{Spontaneous  Electroweak Baryogenesis}}}
\bigskip
\centerline{\it A. G. Cohen
\footnote{$^a$}{Address after 6/18/94:
Department of Physics, Boston University, Boston MA 02215, USA; {\tt cohen@andy.bu.edu}}, 
D. B. Kaplan\footnote{$^b$}
{Address after 6/30/94: Institute for Nuclear Theory HN-12, University of Washington, Seattle WA
98195, USA; {\tt dbkaplan@ben.npl.washington.edu}}, 
A. E. Nelson\footnote{$^b$}
{Address after 6/30/94: Department of Physics FM-15, University of
Washington, Seattle WA 98195, USA; {\tt anelson@galileo.phys.washington.edu}} }
\medskip
\centerline{Institute for Theoretical Physics }
\centerline{University of California}
\centerline{Santa Barbara CA, 93106-4030, USA}
\bigskip
\baselineskip 18pt
\noindent
We include the effects of diffusion in the electroweak spontaneous 
baryogenesis
 scenario and show that it can greatly enhance the resultant baryon density, 
by as much as a factor of $1/\alpha_w^4\sim  10^6$ over previous estimates. 
Furthermore, the baryon density produced is rather insensitive to 
parameters characterizing the first order weak phase transition, such as the 
width and propagation velocity of the phase boundary. 
\Date{6/94}
\newsec{Introduction}
The scenario of electroweak baryogenesis (EWB) \ref\krs {V.A. Kuzmin,    
V.A. Rubakov,
 M.E. Shaposhnikov, \pl{155}{1985}{36}}\ owes its appeal to the fact 
that---unlike most alternatives---it involves particles and interactions 
known to exist, and  may be experimentally testable. (For a relatively recent 
review, see \ref\rev{A.G. Cohen, D.B. Kaplan, A.E. Nelson, Annu. Rev. Part. 
Nucl. Sci. 43 (1993) 27}.) Baryogenesis at the weak scale is made possible 
by the fact that in the symmetric phase at temperature $T$, baryon violation 
is expected to occur at a rate $\Gamma\sim\alpha_w^4 T$, where $\alpha_w$ 
is the weak fine structure constant \ref\hightemp{ P. Arnold,
L. McLerran, 
\physrev{D36}{1987}{581}}. Such theories require a first order 
weak phase transition in order to provide the departure from thermal 
equilibrium necessary for baryogenesis 
\ref\sak{A.D. Sakharov, JETP Lett. 5 (1967) 24.}.  During 
the phase transition, bubbles of  broken $SU_L(2)\times U_Y(1)$ phase nucleate 
and rapidly expand, and the source of nonequilibrium physics 
necessary for baryogenesis is the 
interaction between particles in the plasma and the expanding bubble 
wall that separates the phases.  The greatest challenge for an EWB  
mechanism   is reconciling the fact that nonequilibrium $CP$-violating 
physics is occurring within the bubble wall itself, while baryon 
violation is only rapid outside the bubble in the symmetric phase
(and perhaps in the leading edge of the wall). This difficulty 
is easily surmounted by the  charge transport mechanism 
\nref\neut{A.G. Cohen, D.B. Kaplan, A.E. Nelson, \pl{245}{1990}{561};
 \np{349}{1991}{727}}\nref\some{A.E. Nelson,
D.B. Kaplan, A.G. Cohen, \np{373}{1992}{453}}\nref\debye{
A.G. Cohen, D.B. Kaplan, A.E. Nelson, \pl{294}{1992}{57}}\nref\jptlep{
M. Joyce, T. Prokopec, N. Turok, preprint PUTP-1437 (1994), 
hep-ph/9401352}\refs{\neut-\jptlep}\ in the limit when the bubble 
wall is narrow 
compared to particle mean free paths. With this mechanism  $CP$ 
violation in the phase boundary leads to the creation of net global 
charges among the particles reflecting off the domain wall; the charge 
is then transported into the region in front of the advancing domain wall.
 The global charge distributions in front of the wall contain a nonzero 
density of $SU_L(2)$ doublet fermions,  and therefore bias 
anomalous baryon number 
fluctuations  toward production of baryon number \foot{An 
alternative mechanism  is proposed in \ref\tz{N. Turok, J. Zadrozny,
\prl{65}{1990}{2331}; \np{358}{1991}{471}}, in which nonzero Higgs 
winding number is produced at the bubble wall, subsequently decaying 
in its wake,  producing baryon number.}.   

In the opposite limit, when the width of the bubble wall is large compared 
to the mean free path of fermions in the plasma, the fermions see an 
adiabatically changing background Higgs field  and are not reflected.  Thus 
the charge transport regime does not apply, and alternative analyses have been 
proposed to explain how EWB would proceed within the wall itself.  These 
analyses are based on the idea of ``spontaneous baryogenesis'', where  
a  time varying background $CP$-odd field arising from spontaneous symmetry
 breakdown provides the necessary $CPT$ violation to bias $B$ violating 
interactions toward the creation of net baryon number \ref\sponbarck{A.G. 
Cohen, D.B. Kaplan, \pl{199}{1987}{251};
\np{308}{1988}{913}}.  For EWB it is a nonzero $CP$ violating phase in a 
Higgs field that plays the role of the background  field
and leads to a local excess of $SU_L(2)$ doublet anti-fermions, causing 
anomalous electroweak processes to be biased in favor of producing baryon
 number within the bubble wall  \nref\dhss{M. Dine, P. Huet,
R. Singleton, L. Susskind, \pl{257}{1991}{351}}
\nref\sponbar{A.G. Cohen, D.B. Kaplan, 
A.E. Nelson, \pl{263}{1991}{86}}\nref\susy{A.G. Cohen,  A.E. Nelson, \pl{297}
{1992}{111}}\refs{\dhss-\susy}.
The problem with this scenario is that as the Higgs field turns on,  
baryon violating interactions turn off exponentially fast, the rate being 
proportional to  
$\exp\(-2M_W/T\)$.  Taking the value of the Higgs vev where the baryon 
violation ``cuts off'' to be $\phi_{co}$, it has been argued that 
one can take $\phi_{co}$ to be as large as $\phi_{co}\simeq 14 \alpha_w
T/g$, where 
$g$ is the weak coupling constant \ref\phico{B. Liu, L.  McLerran, N.  
Turok,
\physrev{D46}{1992}{2668}}, in which case the resultant baryon asymmetry 
is proportional to a factor of $\alpha_w^4\simeq 10^{-6}$ 
intrinsic to anomalous baryon violation.  However,  Dine and Thomas
have argued \ref\mdst{M. Dine, S. Thomas, preprint SCIPP-94-01, 
hep-ph/9401265} that in fact $\phi_{co}\simeq \alpha_w T/g$, and that as 
a consequence
the resultant baryon asymmetry is suppressed by an additional  factor 
of $\alpha_w^4$ 
in the two-Higgs model, or $\alpha_w^2$ in the minimal supersymmetric
 model \susy---suppression factors too small to account for the 
observed asymmetry 
$n_B/s\simeq 10^{-10}$ when one includes small CP violating angles.  
The correct value for $\phi_{co}$ is currently not  known, and 
so the Dine--Thomas objection throws in doubt the scenario of 
EWB in the spontaneous baryogenesis  (thick wall) regime. This is 
disturbing, since models of the electroweak sector that do not involve 
$SU_L(2)$ singlet fields generically produce wide domain walls during a 
first order phase transition, and therefore rely on the spontaneous baryogenesis mechanism \nref\walls{M. Dine,
R.L. Leigh,  P. Huet, L. Linde, D. Linde, \pl{283}{1992}{319};
\physrev{46}{1992}{550}}\refs{\walls,\phico}.

In this paper we consider the role of particle diffusion for EWB in the 
spontaneous baryogenesis regime.  The potential importance of diffusion 
was first pointed out by Joyce, Prokopec and Turok \ref\diff{
M. Joyce, T. Prokopec, N. Turok, preprint PUTP-1436, hep-ph/9401351}, 
who concluded that diffusion was inimical to spontaneous baryogenesis.  
In the present work we arrive at a different conclusion---that diffusion 
moves the asymmetry in $SU_L(2)$ doublet fermions from within the bubble wall 
into the  symmetric phase in front, where baryon violation is not exponentially 
suppressed.  As a result the Dine--Thomas suppression factors are evaded, 
and the spontaneous baryogenesis scenario with diffusion ends up looking 
remarkably like the charge transport scenario \refs{\neut,\some}.  

We show that not only can spontaneous electroweak baryogenesis  account 
for the observed baryon asymmetry of the universe, but that the 
asymmetry produced is not very sensitive to the value one takes for 
$\phi_{co}$, 
or to features of the weak phase transition, such as  the width of the 
bubble wall (assuming it to be sufficiently wide), or its velocity. 

By analyzing the dynamical equations associated with spontaneous baryogenesis
we are also able to treat a second important source of suppression 
recently discussed by Giudice and Shaposhnikov 
\ref\giudice{G. Giudice and M. Shaposhnikov, \pl{326}{1994}{118} }.
They pointed out that when all of the quarks are treated as
massless, QCD sphaleron processes equilibrate to zero the
same charge densities that serve to drive electroweak baryogenesis.
We find evidence that QCD and electroweak sphalerons compete,
but that even in the massless quark approximation a nonzero 
baryon density results, which is surprisingly insensitive
to the overall rate of sphaleron processes, being proportional
to the ratio of electroweak to strong sphaleron rates. If this
effect persists in a more sophisticated treatment including finite
quark mass effects then one of the greatest uncertainties about
electroweak baryogenesis --- the overall rate of sphaleron 
processes --- is eliminated.

\newsec{Diffusion equations in the two Higgs doublet model}

To be specific, we will focus on the two Higgs doublet model with 
non-Kobayashi-Maskawa $CP$ violation as a model for electroweak baryogenesis 
\tz, for which the spontaneous baryogenesis mechanism is described in 
\sponbar.  During a first order phase transition bubbles of the broken 
phase will nucleate and expand.  As a top quark in the plasma traverses 
the bubble wall, it interacts with the background Higgs field through its 
 Yukawa coupling $\lambda_t\approx 1$.  In the adiabatic limit of a slowly
moving  
or broad domain wall, the top quark has a spacetime dependent 
mass term of the form $m_t = \lambda_t 
H({\bf r},t)e^{-i\theta({\bf r},t)}$, where 
the spacetime dependent phase $\theta$ arises from $CP$ violation in the 
Higgs potential. The effects of   $\theta$ are most easily analyzed by 
performing a hypercharge rotation of all the fields in the theory, removing 
the $\theta$ dependence from the top quark mass term, at the expense of 
inducing an 
interaction of the form 
\eqn\hyper{-2 \partial_\mu\theta({\bf r},t) J^\mu_Y({\bf r},t)\ .}
With nonzero $\mu\equiv -2\dot\theta$, we see that this interaction 
resembles a thermodynamic charge constraint, with $\mu$ being the 
chemical potential.  Since hypercharge is not conserved inside the domain
 wall, we refer to the dynamically generated $\mu$ as a charge potential, 
to distinguish it from an imposed constraint on a conserved charge. Exploiting 
the resemblance, we know that the
free energy is minimized by having  nonzero particle densities 
$n_i = k_iy_i\mu T^2/6$, where $y_i$ is the hypercharge, 
and $k_i$ a statistical
factor that differs for fermions and bosons.   In 
the limit that all fermions are massless except the top quark, the 
interaction \hyper\ leads to production of $q\equiv(t_L, b_L)$, $t_R$ 
and $H$ particles;
if
$\mu$ is negative, then
 the excess of $q$ doublets (with $Y=+1/6$) is also negative, 
and anomalous electroweak processes 
are biased to produce baryon number \foot{Hypercharge screening 
proves not to  affect the result 
\nref\khleb{S.Yu. Khlebnikov, \pl{300}{1993}{376}}\refs{\khleb, \debye};
 we will return to this point later.}.

To see whether diffusion can be a significant effect, it is useful
to consider the dimensionless quantity 
$$\epsilon_D\equiv D/L_w v_w\ ,$$ 
where $D$ is the diffusion
constant for quarks, $L_w$ is the width of the domain wall, and $v_w$ is the
wall velocity.  None of these quantities is known accurately, 
with estimates for
$v_w$ ranging from $0.1c$ to $c/\sqrt{3}\simeq 0.6c$; $L_w$
ranging from $10/T$ to $40/T$, and $D$ (classically equal to 
 $\ell c/3$, where $\ell$  is the mean free
path) ranging from $1/T$ to $5/T$. 
Evidently $\epsilon_D$ can be $\CO(1)$, and hence one may expect significant (\ie\ $\CO(1)$)
redistributions of particle densities due to diffusion. As we will
show, this eliminates the sensitive dependence on the value of $\phi_{co}$
discussed by Dine and Thomas \mdst. 

In order to understand the detailed effect of the interaction \hyper,
we first derive the diffusion equation relevant for a single particle
species with local number density $j^0 \equiv n$ and number current 
$\vec J$ 
interacting with a slowly varying background
field $A_0$ and $\vec A$ through the interaction 
\eqn\toyint{\CL_{int}=A_\mu j^\mu\ .}
We assume that $j^\mu$ is not exactly conserved, since $j^\mu$ is to
eventually play the role of the hypercharge current, which is violated
in the broken phase. 
Following the spirit of the usual derivation of Fick's law and the
diffusion equation, we perform a simultaneous expansion 
to first order in the deviations from equilibrium $(n-n_0)$,
in derivatives of $(n-n_0)$ and of the background field $A$, and in the
particle number violating interactions.  Keeping only leading terms,
we find
\eqn\curcon{\eqalign{
\dot n &= -\vec\nabla\cdot\vec J[A,n] -{\gamma\over T}{\partial F\over \partial Q}\cr
&= -\vec\nabla\cdot\vec J[A,n] -\Gamma{(n-n_0[A])\over k}\ ,}}
where $F$ is the free energy, $\gamma$ is the violation rate per unit 
volume of the charge $Q=\int d^3\!x\, n$, and we have defined the rate $\Gamma$ as
\eqn\ratedef{\Gamma \equiv {6\gamma\over T^3}\ .}
The parameter $k$ appearing in \curcon\  is  a statistical factor,
\eqn\kdef{
k = {\rm (number\ of\ spin\ degrees\ of\ freedom)}
\times\cases{2&bosons\cr 1&fermions\cr}} up to small
 corrections due to thermal particle masses.
 It remains to find the constituitive relations for $\vec J$ and $n_0$.
Using rotational covariance  we find
\eqn\constit{\eqalign{
\vec J &= -D\vec\nabla n + a_1 \vec A + a_2 \partial_t \vec A +
a_3\vec\nabla A_0 +  \ldots\cr
n_0 &= b_1 A_0 +  \ldots\ ,\cr}}
where the ellipses refer to higher powers of derivatives acting on $n$ and $A$,
as well as terms involving the particle number violating interactions. The 
coefficients $a$ and $b$ are assumed to roughly equal the appropriate
powers of the mean free path $\ell $ dictated by dimensional analysis, so that
the derivative expansion is justified for hydrodynamic modes of
wavelength much longer than $\ell $. (The size of the modes we wish
to consider are set by the bubble wall width $L_w$ assumed to
satisfy $L_w\gg\ell$, which is the spontaneous baryogenesis regime.)

If the current is conserved and $A_\mu$ is  a total divergence, then
the interaction \toyint\ has no physical effect. From this
we deduce that 
$$a_1=0\ ,\qquad a_3 = -a_2\ .$$ It does not follow that 
$b_1=0$,  since this coefficient only appears in eq. \curcon\
proportional to the current nonconserving effect $\Gamma$.  We can
fix $b_1$, however, by considering the simple case  
$A_0=\mu$ and $\vec A=0$.  The minimum of the free energy is then at
$n =k \mu T^2/6$, and so that must be the value $n_0$ 
in eq. \curcon.  It follows that the diffusion equation relevant for the
interaction \toyint\ is
$$ \dot n = D\nabla^2 (n - \phi) -\Gamma {(n-n_0)\over k}\ ,$$
where
$$ D\nabla^2\phi\equiv a_2\vec\nabla\cdot(\partial_t\vec A - \vec\nabla A_0) 
\ ,\qquad n_0 = k A_0 T^2/6\ .$$

This result can be readily taken over to the interaction of interest in
eq. \hyper, and generalized to many particle species.  
Since $A_\mu=-2\partial_\mu\theta$ it follows that $\phi$ in
the above equation vanishes, and that
\eqn\diffeq{
\dot n_i = D_{ij}\nabla^2 n_j - \Gamma_{ij} \({n_j  +   
k_jy_j\dot\theta T^2/ 3\over k_j}\)\ ,}
where $n_i$ is the density of particle species $i$,  $y_i$ is its
hypercharge, and $k_j$ its statistical factor.   
The diffusion matrix $D_{ij}$ has diagonal elements
whose magnitudes are given by the particle mean free paths, while the off-diagonal
elements are smaller by at least one power of $\alpha$, the fine 
structure constant of 
the transition interaction.  We will  work to leading order in perturbation
theory, only keeping the diagonal entries:
$$D_{ij} = D_i\delta_{ij} + \CO(\alpha)\ .$$
As for the $\Gamma_{ij}$---hypercharge is only broken by the vev of
the Higgs field, and so in the limit that we ignore weak mixing
angles, all Yukawa couplings except for the top quark's, and anomalous strong 
interactions,
 the only species of particles whose 
 densities are affected by a
nonzero $\dot\theta$ are the left handed third family doublet
 denoted by $q\equiv (t_L+b_L)$, the right handed top quark $t\equiv t_R$, 
and the Higgs particles $h\equiv (h_1^- + h_1^0 + h_2^- + h_2^0)$.
The individual particle numbers of these
species can change  through the top quark Yukawa interaction, 
the top quark mass, the Higgs self interactions, and 
anomalous weak interactions, at the rates   
$\Gamma_y$, $\Gamma_m$, $\Gamma_h$ and $\Gamma_{ws}$ respectively.

Including strong sphalerons (with a rate $\Gamma_{ss}$)
would allow the generation of right handed bottom
quarks, $b\equiv b_R$, 
as well as first and second family quarks. However since 
strong sphalerons are the only processes which generate 
significant numbers of 
first and second family quarks, and  all quarks  have approximately the same
diffusion constant, these densities are constrained algebraically 
in terms of $b$ to satisfy\eqn\algebra{q_{1L}=q_{2L}=-2u_R=-2d_R=-2s_R=-2c_R=-2b_R\equiv -2b\ .}
To simplify the equations we may use the fact that the only charge potential 
generated is for hypercharge, and thus each $n_0$ is proportional to
$k$ times the hypercharge for the species---this allows us to
eliminate all $n_0$ in favor of  $h_0$. Using the relations
$$
3q_0/k_q + l_0/k_l \propto (3 y_q - y_l) = (1/6)3 - 1/2 =0$$
$$(q_0/k_q -h_0/k_h -t_0/k_t )\propto (y_q - y_h - y_t) = 
(1/6) - (-1/2) - (2/3) =0$$
we see  that the source terms proportional to $\dot\theta$ drop out of $\Gamma_y$,  $\Gamma_{ws}$ interactions and are only present
 proportional to the hypercharge 
violating  $\Gamma_m$ and $\Gamma_h$, consistent with the arguments 
of Dine and Thomas \mdst.

Eq. \diffeq\ may now be written 
as
\eqn\qth{\eqalign{
\dot q &= D_q \nabla^2 q - \Gamma_y\[{q/ k_q} -{h/ k_h} - 
{t/ k_t}\]  -\Gamma_m\[{q/ k_q} -{t/k_t}-h_0/k_h \] \cr
&\qquad -6\Gamma_{ws}\[3{(q-4b)/ k_q} + {l/ k_l}\]  
 - 6 \Gamma_{ss}\[2q/ k_q - t/ k_t
-9 b/k_b\]\cr
\dot t &= D_t \nabla^2 t  - \Gamma_y\[-{q/ k_q} +{h/ k_h} 
+ {t/ k_t}\]  \cr
&\qquad -\Gamma_m\[-{q/ k_q} +{t/ k_t}+ h_0/k_h\] +
3 \Gamma_{ss}\[2q/ k_q - t/ k_t
-9 b/k_b\] \cr
\dot h &= D_h \nabla^2 h - \Gamma_y\[-{q/ k_q} +{t/ k_t}
+ h/k_h\] -\Gamma_h{(h-h_0)/ k_h}\cr
\dot b &= D_b \nabla^2 b +3 \Gamma_{ss}\[2q/ k_q - t/ k_t
-9 b/k_b\] \cr }}
where the $k$ factors are given by \kdef
\eqn\kvals{ k_q=6,\qquad k_t=3,\qquad k_h=8,\qquad k_b=3\ .}
 
Several simplifications of equations \qth\ can be made.  
First we ignore the curvature of the bubble wall, 
and so $\Gamma_m$, $\Gamma_h$,
and $\Gamma_{ws}$
are only functions of $z\equiv\vert\vec r-\vec v_wt\vert$,
where  $\vec v_w$ is the bubble wall velocity.
Explicitly, we assume that
$\Gamma_m$,  and $\Gamma_h$  vary as the square of the Higgs vev,
 while $\Gamma_{ws}$ is approximated by a step function: $\Gamma_{ws}(z)
\simeq \theta(z-z_{co}) \Gamma_{ws}$. The parameters
$D_t$, $D_h$, $\Gamma_y$ and $\Gamma_{ss}$ are taken to be
 spacetime constants.
We will  assume that the density perturbations of interest are 
functions 
of $z$ and $t$ only. 
We will also take 
$D_q\simeq D_t\simeq D_b$, which is correct up to order 
$\alpha_w/\alpha_s$.  
Then, ignoring $\CO(\Gamma_{ws}^2)$ effects on the final baryon density, 
we may set $b=-(q+t)$ and eliminate
the $b$ equation; furthermore, since no leptons are produced in the 
limit that we ignore lepton Yukawa 
couplings we may set $l=0$.

With these assumptions we arrive at the diffusion equations for 
$q(z,t) ,t(z,t)$,  and $h(z,t)$ in
the rest frame of the bubble wall  (where $z$ is the coordinate normal to the surface):

\eqn\masteq{\eqalign{
\dot q &= v_w\partial_z q + D_q \nabla^2 q - \Gamma_y\[q/6 -h/8 - 
t/3\]  -\Gamma_m\[q/6 -t/3-h_0/8 \] \cr
&\qquad -3\Gamma_{ws}\[5q+4t\]   - 2 \Gamma_{ss}\[10 q +8 t\]\cr
\dot t &=v_w\partial_z t  + D_q\partial_z^2 q + \Gamma_y\[q/6 -h/8 - 
t/3\]  \cr
&\qquad  + \Gamma_m\[q/6 -t/3-h_0/8 \]  +\Gamma_{ss}\[10 q +8 t\]\cr
\dot h &=v_w\partial_z h  + D_h\partial_z^2 h + + \Gamma_y\[q/6 -h/8 - 
t/3\] 
-\Gamma_h(z) \[(h-h_0)/8\] \ .\cr}}

In our past treatment of this system \sponbarck\ we  assumed that  
$h$ and $q-t$ (axial top number)  
had time independent solutions in the wall rest 
frame  equal to the values which minimize 
the free energy with zero baryon number. 
This was tantamount to
treating  the dimensionless numbers $\Gamma z_{co}/v_w$ and 
$\Gamma z_{co}^2/D$ as being much larger
than unity.  If $z_{co}$ is very small, as suggested by Dine and Thomas 
\mdst, than this assumption is
false---especially since the rates $\Gamma_m$ and $\Gamma_h$ are 
proportional to the Higgs vev and therefore
vanish at the leading edge of the wall.  Eq. \masteq\ allows us to
deal with realistic values for $\Gamma$, $D$, and $v_w$.

\newsec{Solution of the rate equations}
 
In this paper we
will only use reasonable estimates of the  diffusion 
constants $D$ and rates $\Gamma(z)$ 
in order to demonstrate the
general properties of the solutions to Eq. \masteq.  
We will take the bubble wall 
profile at a transition
temperature $T$  to have  the form 
\eqn\wallshape{\eqalign{ \vev{H(z)}
&\simeq T e^{-i\theta(z)}(1-\tanh(z/L_w))/2 \ ,\cr 
\theta(z)&= \Delta\theta (1-\tanh(z/L_w))/2 \ ,}} 
where $\Delta\theta$ is a measure of the size of $CP$ violation in the theory.
It follows from \diffeq\ that the charge potential $h_0$ 
in eq. \masteq\ is given by
\eqn\equilno{h_0=  - {4\over 3}v_w T^2{{\rm d}\theta\over 
{\rm d}z}}
and that the solutions will scale linearly with $\Delta\theta$.
For the diffusion constants we take 
\eqn\diffcon{D_t = D_q = 3/T\ ,\qquad D_h=10/T\ ,}
where we have used  the estimates from \some\ for $D_q$ and have scaled
$D_h\simeq D_q(\alpha_s/\alpha_w)$.  For the rates (defined
in Eq. \ratedef\ as $\Gamma = 6\gamma/T^3$)  we take
\eqn\equilrates{
\Gamma_h\sim\Gamma_m=\vev{H(z)}^2\lambda_t^2/T\ ,} and
\eqn\sphalrates
{\Gamma_{ws}=6\kappa\alpha_s^4 T\ ,\qquad 
\Gamma_{ss}=6\kappa{8\over3}\alpha_s^4 T\ .}

The steady state  solution to Eqs. 
 \masteq\   was found by numerical integration.
The results for the choice of  parameters 
$\kappa=1$, $v_w=0.1$, $L_w=10/T$ and $\Delta\theta=-\pi$
are given in  \fig\diffigi{
Time independent particle number densities normalized to the 
entropy as a function of position $z$, in the rest frame of the 
expanding bubble wall.  The parameters used were $v_w=0.1$, $L_w=10/T$, $\kappa=1$, and $\Delta\theta=-\pi$; the $D$ and $\Gamma$ 
coefficients used are given in Eqs. \diffcon, \equilrates, \sphalrates.  
The densities $h$, $q-t$, and $q+t$
 refer to Higgs number, $t_L+b_L -t_R$, and $t_L + b_L +t_R$
respectively.  The midpoint of the bubble wall is at $z=0$.}. 
The results clearly indicate that diffusion plays a significant
role, since nonzero particle densities are seen to extend well
beyond the bubble wall into the symmetric phase.  An effect of 
strong sphalerons is seen in the nonzero $q+t$ density, which
vanishes as $\Gamma_{ss}\to 0$.  Solutions to \masteq\ were also obtained 
for a variety of wall velocities ($0.05\le v_w\le 0.3$), wall widths
($10/T\le L_w\le 40/T$), and sphaleron rates ($0.1\le\kappa \le 100$).

The baryon density deep within the bubble  ($z$ large and
negative in these coordinates)
is computed  by simply integrating the equation for
 total lefthanded fermion number
(equal to $(5 q+ 4 t)$, after solving algebraically for the 
number densities
of light quarks) from far outside the bubble down to 
$z_{co}$:
\eqn\bdens{n_B\simeq {3\Gamma_{ws}\over  v_w} \int^\infty_{z_{co}} dz\,
(5q(z)+4t(z))\ .}
As discussed in the introduction, the value of $z_{co}$
is controversial, and a large value (corresponding to small $\phi_{co}$) 
suppresses baryon number when diffusion is ignored.  In
\fig\diffigii{The final baryon to entropy ratio of the universe for several values of $\kappa$, plotted as ${\rm Log}_{10}[(n_B/s)
(g_*/100)(-\pi/\Delta\theta)]$ versus $z_{co}$, the point where weak 
sphalerons become exponentially suppressed.  $z_{co}= -0.5 L_w$ corresponds to
$\phi_{co}\simeq 14\alpha_w T/g$, while $z_{co}=3.0 L_w$ corresponds to $\phi_{co}\simeq \alpha_w T/g$.  The dashed line is the solution to Eq. 
\masteq\ assuming very small diffusion constants, illustrating the 
Dine-Thomas effect.}\ we present the final baryon asymetry $n_B/s$ 
of the universe
as a function of $z_{co}$ for several values of $\kappa$, as
well as the solution for $\kappa=1$ when all the diffusion constants
are taken to be very small, $D_q=D_h=0.1/T$.  Apparently 
spontaneous electroweak baryogenesis can quite easily account for the
observed asymmetry $n_B/s\sim 10^{-10}$ with $CP$ violation at the $\Delta\theta \sim 10^{-2} - 10^{-3}$ level.

There are several remarkable 
features to \diffigii:  First, it illustrates the Dine-Thomas sensitivity 
to the value of $z_{co}$ when diffusion is ignored, but  the sensitivity
is practically eliminated when diffusion is taken into account.  
A second amusing feature is that  
the final baryon asymmetry turns out to be approximately independent 
of $\kappa$ in the range examined.  Ignoring the effects of strong sphalerons
one would have found $n_B\propto \kappa$, but as pointed out by Giudice and Shaposhnikov
\giudice\ 
 the left handed baryon number is driven to zero by strong sphalerons  in
the limit that the thermal fermion masses are neglected.   We find numerically 
that the baryon number with $z_{co}=3.0 L_w$ is inversely proportional to
 $\Gamma_{ss}$ for $0.5\ltap \kappa\ltap 100$, cancelling the linear $\kappa$ dependence for electroweak sphalerons.  For $\kappa\ltap 0.1$, we find $n_B$ to be   insensitive to $\Gamma_{ss}$ and proportional to $\kappa$ (an intuitive understanding of this number is that strong 
sphalerons are irrelevant when the timescale $1/\Gamma_{ss}$ is much 
longer than
the typical time $D/v_w^2$ that the diffusing fermions spend in the symmetric 
phase.) \foot{We have not presented data for $\kappa> 5.0$ since the neglected $m_t$ and $\alpha$ corrections to the relation $n \simeq k\mu T^2/6$ discussed in
\giudice\ are expected to become more important as $\kappa$ gets larger. 
These
effects are expected to further enhance the baryon asymmetry.  Note however 
that even neglecting these effects one finds a sizable baryon asymmetry in our dynamical calculation, unlike in the equilibrium calculation of \giudice.}.

We conclude this section by noting that solutions to Eq. \masteq\ with $\kappa=1$ and a variety of wall velocities and wall widths reveal that the baryon asymmetry one creates is not extremely sensitive to either $v_w$ 
or $L_w$.  We find that for $L_w = 10/T$ and $v_w$ ranging from $0.05$ to $0.3$,
the baryon asymmetry changes by no more than a factor of 2; for a broader 
wall with $v_w=0.1$ and $L_w = 40/T$, $n_B$ was decreased by a factor of 
$\sim 5$.  When we artificially set the diffusion constants 
and $\Gamma_{ss}$ to
be very small and assume $\phi_{co}=14\alpha_w T/g$, $\kappa=1$,
 we reproduce our earlier results 
 which neglected both diffusion and 
strong sphalerons \foot{Our review \rev\ contains a typographical
 error in Eq. 37:  the right side of Eq. 37 should 
include the factor $({\cal N}/.33)$ instead of $({\cal N}/.1)$}.

\newsec{Outlook}

We conclude that a dynamical treatment of  spontaneous electroweak 
baryogenesis with finite interaction rates, nonzero diffusion 
constants, and strong sphaleron effects eliminates or moderates large suppression factors found in an equilibrium calculation.  Our work indicates that the EWB scenario remains a viable explanation for the baryon asymmetry of the universe.  Furthermore, the
results appear not to be very sensitive to the bubble wall velocity or 
width, to the point in the wall where anomalous baryon violation becomes suppressed, or to the sphaleron rate parameter $\kappa$ for a broad 
range of these parameters.
This is encouraging, since these quantities are poorly known even in
specific models.
On the other hand, our results are rather sensitive  to the values of the diffusion lengths and interaction rates, which we have only estimated.  These
quantities are computable from known physics, and this should be done before making firm conclusions.  Corrections of order $m_t^2$ and $\alpha$ to the 
relationship between density and charge potential should also be included,
particularly for large $\kappa$.  

A more complete treatment would also include the effects of hypercharge screening \refs{\khleb,\debye}.  Screening is known not to affect the spontaneous baryogenesis in equilibrium calculations, since three quark 
doublets are drawn in for every anti-lepton doublet, so that there is no 
net effect on weak sphalerons.  In a dynamical calculation one might expect 
the anti-leptons to screen more efficiently since they have a longer mean 
free path, in which case the baryon number produced would only be enhanced.

Finally, we note that diffusion will be even more important for 
spontaneous baryogenesis in the minimal supersymmetric model \susy, 
due to the contributions from charginos and neutralinos which have 
much longer mean free paths than quarks.  In addition, strong 
sphaleron suppression will be absent due the combinations of charges 
that bias baryon number in that model.

\vfill
\centerline{Acknowledgements}
We wish to thank 
the ITP for hospitality during the workshop Weak Interactions '94 
where this work was completed, and supported in part under the 
NSF grant PHY89-04035.
A.G.C. was supported in part under grant DE-FG02-91ER40676; 
D.B.K. and A.E.N. were supported in part by funds from the DOE.

\listrefs
\listfigs
\vfill\eject
\centerline{\epsfbox{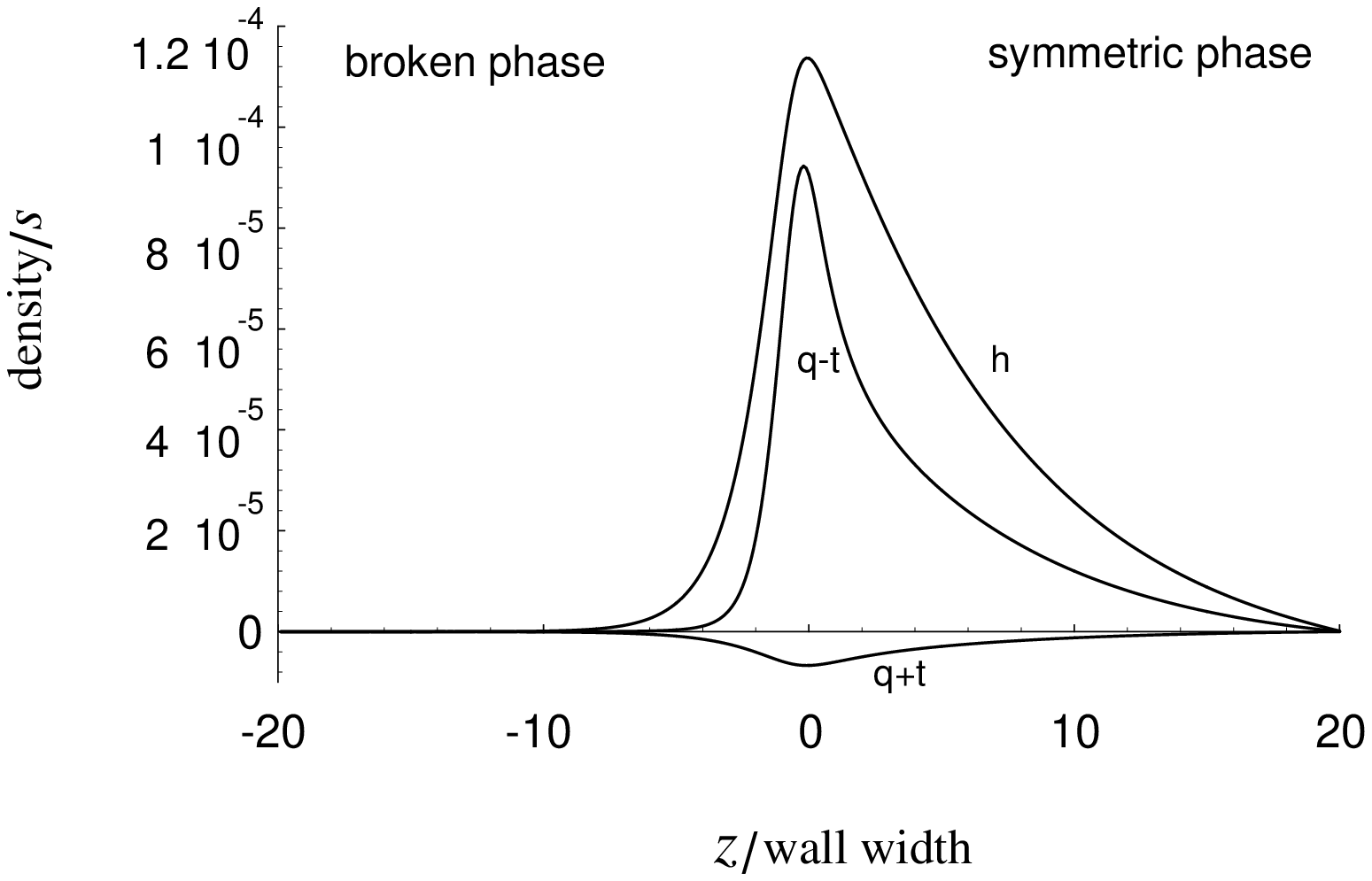}}
\vskip 1in
\centerline{\titlefont Figure 1}
\vfill\eject
\centerline{\epsfbox{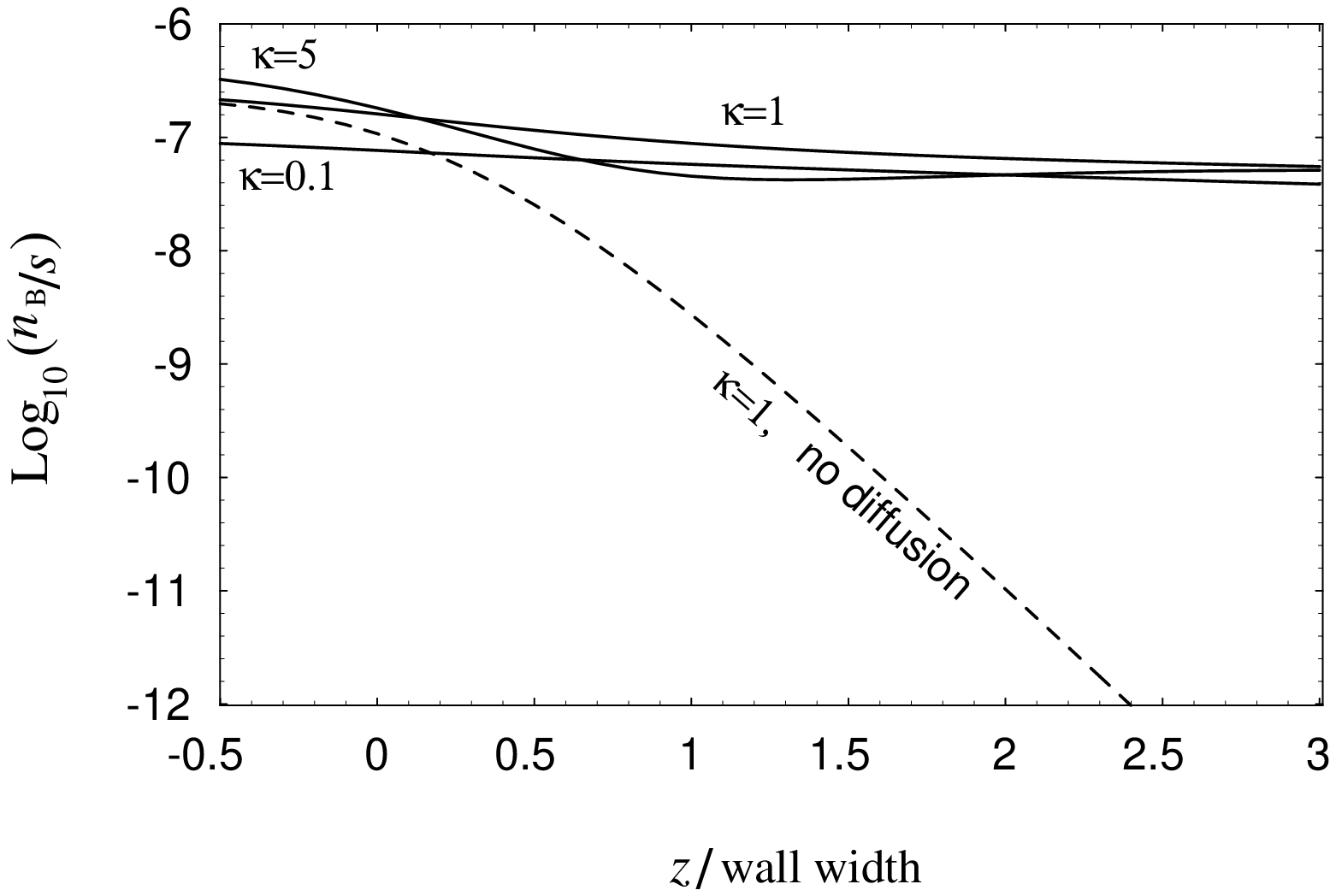}}
\vskip 1in
\centerline{\titlefont Figure 2}
\vfill\eject
\bye